\documentclass[12pt]{article}
\begin{document}

\title{\bf \Large Force Behaviour in Radiation-Dominated Friedmann Universe}

\author{M. Sharif \thanks{e-mail: hasharif@yahoo.com}
\\ Department of Mathematics, University of the Punjab,\\ Quaid-e-Azam
Campus Lahore-54590, PAKISTAN.}

\date{}

\maketitle

\begin{abstract}
We consider radiation-dominated Friedmann universe and evaluate
its force four-vector and momentum. We analyse and compare the
results with the already evaluated for the matter-dominated
Friedmann model. It turns out that the results are physically
acceptable.
\end{abstract}

\section{Introduction}

It is of cosmological interest to consider a fluid composed of
incoherent radiations. The primordial radiation played a dominant
role in the early universe for the times smaller than the times of
re-combination. We have obtained new insights by considering the
force four-vector [1] and momentum [2] for the matter-dominated
Friedmann universe. Further insights can be expected by analysing
the same quantities for the radiation-dominated Friedmann
universe. The procedure adopted for this purpose is the extended
pseudo-Newtonian $(e\psi N)$-formalism [1].

Einstein's Theory of Relativity replaces the use of forces in
dynamics by what Wheeler calls ``geometrodynamics''. Paths are
bent, not by forces, but by the ``curvature of spacetime''. In the
process the guidance of the intuition based on the earlier
dynamics is lost. However, our intuition continues to reside in
the force concept, particularly when we have to include other
forces in the discussion. For this reason one may want to reverse
the procedure of General Relativity (GR) and look at the
non-linear ``force of gravity'' which would predict the same
bending of the path as predicted by geometry. In the $\psi
N$-approach [3] the curvature of the spacetime is ``straightened
out'' to yield a relativistic force which bends the path, so as to
again supply the guidance of the earlier, force-based, intuition.

The idea of re-introducing the Newtonian gravitational force into
the theory of GR arose in an attempt to deal with the following
problem: Gravitation, being non-linear, should dominate over the
Coulomb interaction at some, sufficiently small, scale. At what
scale would it occur? Whereas this question is perfectly valid in
pre-relativistic terms it becomes meaningless in GR. The reason is
that gravitation is expressed in purely geometric terms while
electromagnetism is not. Thus, in Relativity, gravitation
possesses a very different status than the other forces of Nature.
Our physical intuition for the other interactions, nevertheless,
rests on the concept of forces. To deal with gravity and other
forces together, we must either express the other forces
geometrically, as in the Kaluza-Klein theories, or express
gravitation in the same terms as the other forces. We will follow
the latter alternative as the simpler program to implement.

The procedure adopted converts an idealised operational definition
of the gravitational force (via the tidal force) into a
mathematical formulation. The relativistic analogue of the
Newtonian gravitational force which gives the relativistic
expression for the tidal force in terms of the curvature tensor is
called the $e\psi N$-force. The quantity whose proper time
derivative gives the $e\psi N$-force is the four-vector momentum
of a test particle. The spatial components of this vector give the
momentum imparted to a test particle as defined in the preferred
frame $(g_{0i}=0=g_{ab,0i}) (a,b=0,1,2,3)$. The plan of the paper
is as follows: In the next section we shall give a brief
description of the formalism employed. In section three, we shall
apply the formalism to evaluate force four-vector and momentum
four-vector for radiation-dominated Friedmann models. In the last
section we shall conclude the results achieved.

\section{The $e\psi N$-Formalism}

The basis of the formalism is the observation that the tidal
force, which is operationally determinable, can be related to the
curvature tensor by

\begin{equation}
F_T^\mu=mR_{\nu\rho\pi}^\mu t^\nu l^\rho t^\pi,\quad
(\mu,\nu,\rho,\pi=0,1,2,3),
\end{equation}
where $m$ is the mass of a test particle, $t^\mu=f(x)\delta_0^\mu,
\quad f(x)=(g_{00})^{-1/2}$ and $l^\mu$ is the separation vector.
$l^\mu$ can be determined by the requirement that the tidal force
have maximum magnitude in the direction of the separation vector.
Choosing a gauge in which $g_{0i}=0$ (similar to the synchronous
coordinate system [4]) in a coordinate basis. We further use
Riemann normal coordinates (RNCs) for the spatial direction, but
not for the temporal direction. The reason for this difference is
that both ends of the accelerometer are spatially free, i.e., both
move and do not stay attached to any spatial point. However, there
is a ``memory" of the initial time built into the accelerometer in
that the zero position is fixed then. Any change is registered
that way. Thus ``time" behaves very differently from ``space".

The relativistic analogue of the Newtonian gravitational force
called the $\psi N$ gravitational force, is defined as the
quantity whose directional derivative along the accelerometer,
placed along the principal direction, gives the extremised tidal
force and which is zero in the Minkowski space. Thus the $e\psi N$
force, $F_\mu$, satisfies the equation

\begin{equation}
F_T^{*\mu}=l^\nu F_{;\nu}^\mu ,
\end{equation}
where $F_T^{*\mu}$ is the extremal tidal force corresponding to
the maximum magnitude reading on the dial. Notice that
$F_T^{*0}=0$ does not imply that $F^0=0$. With the appropriate
gauge choice and using RNCs spatially, Eq.(2) can be written in a
space and time break up as

\begin{equation}
l^i(F_{,i}^0+\Gamma_{ij}^0F^j)=0,
\end{equation}
\begin{equation}
l^j(F_{,j}^i+\Gamma_{0j}^iF^0)=F_T^{*i}.
\end{equation}

A simultaneous solution of the above equations can be found by
taking the ansatz

\begin{equation}
F_0=-m\left[\{\ln
(Af)\}_{,0}+g^{ik}g_{jk,0}g^{jl}g_{il,0}/4A\right], \quad
F_i=m(\ln f)_{,i},
\end{equation}
where $A=(\ln \sqrt{-g})_{_{,0}},\quad g=det(g_{_{ij}})$. This
force formula depends on the choice of frame, which is not
uniquely fixed.

The new feature of the $e\psi N$ force is its zero component. In
special relativistic terms, which are relevant for discussing
forces in a Minkowski space, the zero component of the four-vector
force corresponds to a proper rate of change of energy of the test
particle. Further, we know that in general an accelerated particle
either radiates or absorbs energy according as $\frac{dE}{dt}$ is
less or greater than zero. Thus $F_0$, here, should also
correspond to energy absorption or emission by the background
spacetime. Infact we could have separately anticipated that there
should be energy non-conservation as there is no timelike
isometry. In that sense $F_0$ gives a measure of the extent to
which the spacetime lacks isometry.

Another way of interpreting $F_0$ is that it gives measure of the
change of the "gravitational potential energy" in the spacetime.
In classical terms, neglecting this component of the $e\psi N$
force would lead to erroneous conclusions regarding the "energy
content" of the gravitational field. Contrariwise, including it
enables us to revert to classical concepts while dealing with a
general relativistically valid treatment. It can be hoped that
this way of looking at energy in relativity might provide a
pointer to the solution of the problem of definition of mass and
energy in GR.

The spatial component of the $e\psi N$ force $F_i$ is the
generalisation of the force which gives the usual Newtonian force
for the Schwarzschild metric and a $"\frac{Q^2}{r^3}"$ correction
to it in the Riessner-Nordstrom metric. The $\psi N$ force may be
regarded as the "Newtonian fiction" which "explains" the same
motion (geodesic) as the "Einsteinian reality" of the curved
spacetime does. We can, thus, translate back to Newtonian terms
and concepts where our intuition may be able to lead us to ask,
and answer, questions that may not have occurred to us in
relativistic terms. Notice that $F_i$ does not mean deviation from
geodesic motion.

The quantity whose proper time derivative is $F_\mu$ gives the
momentum four-vector for the test particle. Thus the momentum
four-vector, $p_\mu$, is [5]

\begin{equation}
p_\mu=\int F_\mu dt.
\end{equation}
The spatial components of this vector give the momentum imparted
to test particles as defined in the preferred frame (in which
$g_{_{0i}}=0)$.

\section{Radiation-Dominated Friedmann Universe}

The metric describing the standard model [6,7] is the Friedmann
metric

\begin{equation}
ds^2=dt^2-[\frac {dr^2}{1+kr^2/a^2(t)}+r^2 d \Omega^2],
\end{equation}
where $k$ can take the values $-1, 0$ or 1. In these three cases a
new variable, $\chi$, can be used in place of $r$
\begin{equation}
open (k=-1),\quad r=a(t)\sinh\chi
\end{equation}
\begin{equation}
flat (k=0),\quad r=a(t)\chi
\end{equation}
\begin{equation}
closed (k=+1),\quad r=a(t)\sin\chi
\end{equation}
to give the metric
\begin{equation}
ds^2=dt^2-a^2(t)[d\chi^2+\sigma^2(\chi)d\Omega^2],
\end{equation}
where $\chi$ is the hyperspherical angle, $\sigma(\chi)=\sinh\chi,
\chi, \sin\chi$ as for $k=-1,0,1$ and $a(t)$ is a scale parameter.
For radiation-dominated universe, it is given [8] by

\begin{equation}
a(t)=a_1\sinh\eta,\quad t= a_1(\cosh\eta-1)(0\le\eta<\infty),\quad
k=-1
\end{equation}
\begin{equation}
a(t)=\sqrt{2a_1t},\quad k=0
\end{equation}
\begin{equation}
a(t)=a_1\sin\eta,\quad t=a_1(1-\cos\eta) (0\le\eta\le 2\pi),\quad
k=+1
\end{equation}
or by [9]
\begin{equation}
a(t)=\sqrt{2a_1t+t^2}\quad (k=-1),\quad \sqrt{2a_1t}\quad
(k=0),\quad \sqrt{2a_1t-t^2}\quad (k=+1)
\end{equation}

The $e \psi N$-force, for the Friedmann models, is given [1] as
\begin{equation}
F_0=-m\frac{\ddot a}{\dot a},\quad F_i=0,
\end{equation}
where dot denotes differentiation with respect to the coordinate
time t. The corresponding $p_0$ and momentum $p_i$, imparted to
test particle, are given [2] by

\begin{equation}
p_0=-m\ln(c \dot a),\quad p_i=constant,
\end{equation}
where c is an integration constant.

Now we evaluate the $e\psi N$-force and momentum for all the three
cases of radiation-dominated Friedmann models.

For the flat model, it will be

\begin{equation}
F_0=\frac{m}{2t},\quad F_i=0
\end{equation}
Thus $F_0$ is proportional to $t^{-1}$ and hence $F_0$ goes to
infinity as t approaches to zero and it tends to zero when t tends
to infinity. Since $F_0$ is positive, it corresponds to the energy
absorption [1] by the background spacetime. The corresponding
$p_0$ and $p_i$ will become

\begin{equation}
p_0=m\ln (\frac{\sqrt{2t}}{c\sqrt{a_1}}),\quad p_i=constant
\end{equation}

For the open Friedmann model, Eq.(5) gives

\begin{equation}
F_0=-\frac{m}{a_1sinh^2\eta cosh\eta},\quad F_i=0
\end{equation}
Hence $F_0$ goes as $t^{-3}$ for large values of t. Consequently,
the quantities $p_0$ and $p_i$ turn out to be

\begin{equation}
p_0=m\ln(\frac{\tanh\eta}{c}),\quad p_i=constant.
\end{equation}

For the closed Friedmann universe, it will be
\begin{equation}
F_0= \frac{m}{a_1sin^2\eta cos\eta},\quad F_i=0
\end{equation}
Thus $F_0$ turns out to be infinite for $\eta=0,\pi,2\pi$. The
corresponding $p_0$ and $p_i$

\begin{equation}
p_0=m\ln(\frac{tan\eta}{c}),\quad p_i=constant.
\end{equation}

\section{Conclusion}

It has been shown that the $e\psi N$-formalism, which had been
useful in providing insights for matter-dominated Friedmann
universe, can be used for radiation-dominated Friedmann universe.
Further, we have evaluated the momentum 4-vector. It is shown that
the behaviour of flat Friedmann universe is same for both matter
and radiation-dominated universes. The temporal component of force
goes to zero for large values of t. However, the spatial
components are zero in all the three cases as there is no
gravitational source present in the Friedmann universe. For the
open and closed radiation-dominated universes, the behaviour of
the temporal component of the force is similar and it goes to
infinity as $\eta$ tends to 0. It is worth mentioning here that
the force does not remain finite even for closed model at
$\eta=\pi$ though it was finite for the matter-dominated universe
[1]. This is the difference between the matter and
radiation-dominated models. The reason is that radiation-dominated
era is the early universe and at that time everything is same and
there is no concept of closedness. Thus one justifies the infinite
rate of change of energy at the early universe.

The momentum 4-vector shows that the temporal component, which
gives energy of a test particle, is infinite in all the three
cases. However, the momentum of a test particle turns out to be
constant in each case and this exactly coincides with the
matter-dominated model already evaluated [2].

\vspace{2cm}


\begin{description}
\item  {\bf Acknowledgment}
\end{description}

I am grateful to Prof. Chul H. Lee for his hospitality at the
Department of Physics and Korea Science and Engineering Foundation
for postdoc Fellowship at Hanyang University Seoul, Korea. A
partial help of BK21 program is also appreciated.

\vspace{2cm}

{\bf \large References}

\begin{description}

\item{[1]} Qadir, Asghar and Sharif, M. Nuovo Cimento B {\bf
107}(1992)1071;\\ Sharif, M. Ph.D. Thesis Quaid-i-Azam University
Islamabad (1991).

\item{[2]} Sharif, M. ``Momentum and Angular Momentum in the
Expanding Universe'' Astrophysics and Space Science {\bf
262}(1999)297.

\item{[3]} Qadir, A. and Quamar, J. {\it Proc. third Marcel
Grossmann Meeting on General Relativity}, ed. Hu Ning (North
Holland Amstderm 1983)189;\\ Qaumar, J. Ph.D. Thesis Quaid-i-Azam
University Islamabad (1984).

\item{[4]} Landau, L.D. and Lifschitz, E.M. {\it
The Classical Theory of Fields} (Pergamon Press 1975).

\item{[5]} Qadir, Asghar and Sharif, M. Phys. Lett. A {\bf
167}(1992)331.

\item{[6]} Misner, C.W., Thorne, K.S. and Wheeler, J.A. {\it
Gravitation} (W.H. Freeman San Francisco 1973).

\item{[7]} Peebles, P.J.E. {\it Principles of Physical Cosmology}
(Princeton University Press 1993);\\ Kolb, E.W. and Turner, M.S.
{\it The Early Universe} (Addison-Wesley 1990).

\item{[8]} Grib, A. Andrey {\it Early Expanding Universe and
Elementary Particles} (Friedmann Laboratory Pub. Ltd. 1995).

\item{[9]} Raychaudhuri, A.K., Banerji, S. and Banerjee, A. {\it
General Relativity and Cosmology} (Springer-Verlag New York Inc.
1992)

\end{description}

\end{document}